\documentclass[onecolumn]{aastex6}

\fullcollaborationName{The Friends of AASTeX Collaboration}

\begin{document}


\title{The Effects of Physically Unrelated
Near Neighbors on the Galaxy-Galaxy Lensing Signal}



\author{Tereasa G. Brainerd\altaffilmark{1}}

\affil{Department of Astronomy \\
Boston University \\
725 Commonwealth Avenue \\
Boston, MA 02215 USA}

\altaffiltext{1}{brainerd@bu.edu}

\begin{abstract}


The effects of near neighbors on
the galaxy-galaxy lensing signal are investigated using a suite of Monte Carlo
simulations.  The redshifts, 
luminosities, and relative coordinates for the simulated lenses were
obtained from a set of galaxies with known spectroscopic redshifts
and known luminosities.  
As expected,
when all lenses are assigned a single, fixed redshift,
the mean tangential shear is identically equal to
the excess surface mass density, scaled by the critical surface
mass density: $\gamma_T = \Delta\Sigma \times \Sigma_c^{-1}$.  
When the lenses are assigned their observed redshifts and
$\Sigma_c$ is taken to be the critical surface mass density of the central lens,
the relationship $\gamma_T = \Delta\Sigma \times \Sigma_c^{-1}$ is violated
because $\gtrsim 90$\% of the near neighbors 
are located at redshifts significantly different from the
central lenses.  
For a given
central lens,
physically unrelated near neighbors give rise
to a ratio of $\gamma_T$ to $\Delta\Sigma \times \Sigma_c^{-1}$
that spans a wide range of $\sim 0.5$ to $\sim 1.5$ at projected distances
$r_p \sim 1$~Mpc.  The
magnitude and sense of the discrepancy between $\gamma_T$ and $\Delta\Sigma
\times \Sigma_c^{-1}$ are functions of both $r_p$ and the velocity dispersions
of the central lenses, $\sigma_v$.  At large $r_p$, the difference between
$\gamma_T$ and $\Delta\Sigma \times \Sigma_c^{-1}$ is, on average,
much greater for low-$\sigma_v$
central lenses than it is for high-$\sigma_v$
central lenses.

\end{abstract}



\section{Introduction} 

Galaxy galaxy lensing, the systematic weak lensing of background galaxies
by foreground galaxies, is powerful method by which the amount of dark matter
and its relationship to luminous matter can be directly constrained.  The
first statistically-significant ($\gtrsim 4\sigma$) detections of galaxy-galaxy lensing to be
published in the peer-reviewed literature demonstrated the viability of
galaxy-galaxy lensing as a cosmological tool, but the sample sizes of 
foreground and background galaxies were too small to place particularly 
strong constraints on the properties of the dark matter halos of the
foreground galaxies (e.g., Brainerd et al.\ 1996;
Dell'Antonio \& Tyson 1996; Griffiths et al.\ 1996; Hudson et al.\ 1998).  
Later studies were
able to take advantage of increasingly large samples, leading to improved
constraints and demonstrating the importance of large surveys in the 
detection of the galaxy-galaxy lensing signal
(e.g., Fischer et al.\ 2000;
Guzik \& Seljak 2002; Hoekstra et al.\ 2004, 2005; Sheldon et al.\ 2004;
Mandelbaum et al.\ 2005, 2006; Heymans et al\ 2006; Kleinheinrich et al.\ 2006;
Tian et al.\ 2009).
More recently, increasingly large data sets from the Sloan Digital Sky 
Survey (York et al.\ 2000), the Dark Energy Survey
(Flauger 2005), the COSMOS survey (Koekemoer et al.\ 2007;
Scoville et al.\ 2007ab),
the Red Sequence Cluster Survey 2 (Gilbank et al.\ 2011), the Canada-France-Hawaii
Telescope Lensing Survey (Heymans et al.\ 2012; Erben et al. 2013), and the combination
of the Galaxy And Mass Assembly Survey (Driver et al.\ 2009, 2011; Liske et al.\
2015) with the Kilo Degree Survey (Kuijken et al.\ 2015)
have continued to
dramatically improve the constraints that can be placed on the relationship
between dark and luminous matter from galaxy-galaxy lensing.  
These most recent
studies have placed strong constraints on the stellar-to-halo mass relation,
the luminosity-to-halo mass relation, 
the dependence of the average halo mass on cosmic environment, and
the masses of the halos surrounding passive versus star-forming galaxies
(e.g., van Uitert et al.\ 2011, 2015, 2016;
Leauthaud 2012ab; Tinker et al. 2013; 
Brimioulle et al.\ 2013;
Velander et al.\ 2014;
Coupon et al.\ 2015; Hudson et al.\ 2015; Zu \& Mandelbaum 2015;
Brouwer et al.\ 2016; Clampitt et al.\ 2016; Mandelbaum et al.\ 2016).

A key goal for galaxy-galaxy lensing studies is a measurement
of the surface mass density surrounding central lens galaxies via observations
of the mean tangential shear, $\gamma_T(r_p)$. In the case
of isolated, axisymmetric lenses, it is well-known that the mean tangential
shear is related to the excess surface mass density,
$\Delta\Sigma(r_p)$, through
\begin{equation}
\Delta\Sigma(r_p) \equiv \left< \Sigma(< r_p) \right> - \Sigma(r_p) =
\Sigma_c ~ \gamma_T(r_p)
\label{eq:sigma_gamma}
\end{equation}
where $\left< \Sigma(< r_p) \right>$ is the mean interior surface mass density contained
within a circle of radius $r_p$, centered on the lens, and
$\Sigma(r_p)$ is the surface mass density at radius $r_p$ from
the lens (e.g.,
Miralda-Escud\'e 1991).  The quantity $\Sigma_c$ is known as the 
critical surface mass density and is given by
\begin{equation}
\Sigma_c = \frac{c^2}{4\pi G} \frac{D_s}{D_l D_{ls}}
\label{eq:sigma_c}
\end{equation}
where $G$ is Newton's gravitational constant,
$c$ is the velocity of light, $D_l$ is the angular diameter distance between the
observer and the lens, $D_s$ is the angular diameter distance between the observer
and the source, and $D_{ls}$ is the angular diameter distance between the lens and the 
source.  Equation~\ref{eq:sigma_gamma} is also true in the case
of an isolated non-axisymmetric lens so long as $\gamma_T(r_p)$ and
$\Sigma(r_p)$ are interpreted as mean 
values, averaged over a ring of radius $r_p$, centered
on the lens (see, e.g., Schneider 2006).  

In the case of multiple lens galaxies
that all share the same redshift (as would be the case for, say, the halos of satellite galaxies
that are contained within a larger system), Equation~\ref{eq:sigma_gamma} is also
strictly true.  It is this latter property of Equation~\ref{eq:sigma_gamma} that is
commonly used to convert an observed value of the mean tangential shear into a constraint
on the excess surface mass density in galaxy-galaxy lensing studies.  
Since the value of $\Sigma_c$ for a given lens-source
pair depends upon the redshifts of both the lens and the source, the conversion of
the shear into a constraint on the excess surface mass density requires a computation
of the value of $\Sigma_c$ that is appropriate for the sample.  This is sometimes done 
by computing a mean value of $\Sigma_c$ for the entire sample, and the
mean is computed over the full lens-source distribution in redshift space. In that
case,  
the right hand side of Equation~\ref{eq:sigma_gamma} is evaluated as the product of the
mean tangential shear, computed over all lens-source pairs, and the mean value of
$\Sigma_c$.  More often, a value of $\Sigma_c$ is computed for each 
lens-source pair, then the products of the individual values of $\Sigma_c$ and $\gamma_T$
for each lens-source pair are averaged together to infer the mean excess surface mass
density.  Note that, in practice, various weighting schemes are typically
used in observational studies in order to optimize the detection of the weak lensing
shear and the reader is referred to recent papers on this subject for examples
of these weighting schemes (see, e.g.,
Velander et al.\ 2014; Hudson et al.\ 2015; Brouwer et al.\ 2016).

The published observational constraints on $\Delta\Sigma(r_p)$ have all acknowledged that,
in addition to shear caused by the central lens galaxies around which $\gamma_T(r_p)$
is computed (the so-called ``one-halo'' term), there is an additional contribution to 
$\gamma_T(r_p)$
at large $r_p$ due to neighboring galaxies
(the so-called ``two-halo'' term).  For the most part, the two-halo term has been interpreted
as being caused by neighboring galaxies that are physically associated with the
central lens galaxies.  That is, it is generally assumed that the two-halo term is dominated
by neighboring galaxies that share the same redshift as the central lens galaxies (see, e.g.,
the review by 
Mandelbaum 2015).  In the case that the two-halo term is, indeed, dominated by physically
associated near neighbor galaxies, then  
Equation~\ref{eq:sigma_gamma} can be used 
to infer the excess surface mass density directly from the observed mean tangential shear, 
with $\Sigma_c$ being the value of the critical surface mass density that is
appropriate for a given central
lens, its physically associated near neighbors, and a given background source.

It is, however, not necessarily the case that the two-halo term is dominated by physically
associated near neighbors.  The first simulations of galaxy-galaxy lensing showed that,
for a given lens galaxy, the closest lens on the sky was not necessarily the most important
lens and, typically, a given source was lensed at a comparable 
level by at least 3 or 4 physically unrelated
foreground galaxies
(Brainerd et al.\ 1996).  
Additionally, Brainerd (2010) showed that, for a realistic population
of lens galaxies with median redshift $z_{\rm med} = 0.55 $, a population of source galaxies with
median redshift $z_{\rm med} = 0.96$ would 
experience a significant amount of weak lensing
due to multiple, physically unassociated galaxies.

The goal of this paper is to examine the degree to which physically unassociated near
neighbors affect the galaxy-galaxy lensing signal in
the limit of a realistic, relatively deep 
sample of lens galaxies.  Using a set of galaxies with known
spectroscopic redshifts and known luminosities, a suite of Monte Carlo
simulations of galaxy-galaxy lensing were constructed using a simple, analytic
dark matter halo model.  From the simulations, the relationship between the mean
tangential shear and the actual surface mass density can be investigated directly, and
the net effects of multiple, physically unassociated lens galaxies on the galaxy-galaxy
lensing signal can be assessed straightforwardly.

The paper is organized as follows.  Details of the galaxy-galaxy
lensing simulations are discussed in Section~2, and results for the
mean tangential shear and the excess surface mass density, scaled
by the critical surface mass density,
are presented
in Section~3.  A summary and discussion of the results are presented
in Section~4.  Throughout, the present-day values of the cosmological
parameters are taken to be $\Omega_0 = 0.25$, $\Omega_{\Lambda 0} = 0.75$,
and $H_0 = 70$~km~sec$^{-1}$~Mpc$^{-1}$.

\section{Simulations of Galaxy-Galaxy Lensing} 

To ensure the greatest realism in the simulations, 
relatively bright galaxies with known redshifts
and known rest-frame blue luminosities ($L_B$) were selected from
a region of diameter 4$'$, centered on the Hubble Deep Field-North
(HDF-N; Williams et al.\ 1996).  
This region of sky was the subject of both a deep redshift survey
(Cohen et al.\ 1996; Steidel et al.\ 1996; Lowenthal et al.\ 1997;
Phillips et al.\ 1997; Cohen et al.\ 2000) and an extensive multicolor
photometric investigation (Hogg et al.\ 2000).  For this investigation,
the redshifts of the galaxies were obtained from Tables~2A and B from Cohen et al.\ (2000) and
their rest-frame blue luminosities were obtained from Table~1 of Cohen (2001).

The completeness limits of the redshift survey differ for the 
HDF-N itself and the regions that surround it, with the completeness limit
being shallower in the surrounding regions than it is in the HDF-N.  
In order assess the effects of near
neighbors on the galaxy-galaxy lensing signal, however, it is important
that the redshift distributions of the lenses be the same within
both the HDF-N and the surrounding regions.  Therefore, 
for this investigation a conservative
completeness limit of $R = 23$ was imposed, resulting in the selection of
427 galaxies from Cohen et al.\ (2000) and Cohen (2001).
The observed celestial coordinates, redshifts,
and rest-frame blue luminosities of these galaxies were then used as the basis of 
a suite galaxy-galaxy lensing simulations in which the effects of differing
numbers of near neighbors on the net mean tangential shear
could be assessed directly.
The advantage to using these particular galaxies
for the simulations is that the relative lensing strengths of the
galaxies are effectively known, since both the redshifts 
and the relative depths of the dark matter potential wells of the galaxies are known.

This investigation is not focussed on an accurate reproduction of an 
observed galaxy-galaxy lensing signal (i.e., such as one would want
for the purpose of 
constraining the physical properties
of the dark matter halos of the lens galaxies).  Instead, here
the focus is a demonstration of the net effect of near neighbor
galaxies on the galaxy-galaxy lensing signal.
Because of this, the simulations do not
incorporate the contribution of all galaxies within this region of space to the
galaxy-galaxy lensing signal.  Rather,
the simulations are restricted to the effects of, at most, the four nearest neighbors.
The angular separations between each central lens galaxy
and its near neighbors are denoted by $\theta_1$, $\theta_2$, $\theta_3$, and
$\theta_4$, where $\theta_1$ is the separation between a central lens galaxy
and its nearest neighbor, $\theta_2$ is the separation between a central 
lens galaxy and its second nearest neighbor, etc.  In order to avoid edge effects
that would be caused by not including weak lensing due to galaxies that are located
just outside the region of the redshift survey, only those galaxies that are
farther than $(4' - \theta_4)$ from the edge of the survey are selected as
central lenses.  A total of 348 galaxies with $R \le 23$ are located
sufficiently far from the edge of the survey to be selected as central 
lenses.  

For simplicity, the dark matter halos of all galaxies were modeled as singular
isothermal spheres, for which the surface mass density is given by
\begin{equation}
\Sigma(\theta) = \frac{\sigma_v^2}{2G D_l \theta} .
\label{eq:sigma_sis}
\end{equation}
The Einstein radius of a singular isothermal lens is given by
\begin{equation}
\theta_E = 4\pi \left( \frac{\sigma_v}{c} \right)^2 \frac{D_{ls}}{D_s}, 
\end{equation}
and the shear for a singular isothermal lens is given by
\begin{equation}
\gamma(\theta) = \frac{\theta_E}{2\theta} .
\end{equation}
Here 
$\sigma_v$ is the velocity dispersion of the dark matter halo surrounding
the lens galaxy and
$\theta$ is the angular separation on the sky between the lens and the source.
While this halo mass distribution is unphysical,
it has the advantage of being both analytic and simple, allowing the results
presented here to be reproduced both easily and straightforwardly.

Velocity dispersions for the halos of the galaxies
were assigned using a simple Tully-Fisher or Faber-Jackson type of
relation:
\begin{equation}
\frac{\sigma_v}{\sigma_v^\ast} = \left( \frac{L_B}{L_B^\ast}  \right)^{1/4}
\end{equation}
where $\sigma_v^\ast$ is the velocity dispersion of the halo of a galaxy
with rest-frame blue luminosity $L_B^\ast$.  For the simulations 
presented here, a value of $\sigma_v^\ast = 156$~km~sec$^{-1}$
was adopted. 


\begin{figure}[t!]
\figurenum{1}
\centerline{\scalebox{0.90}{\includegraphics{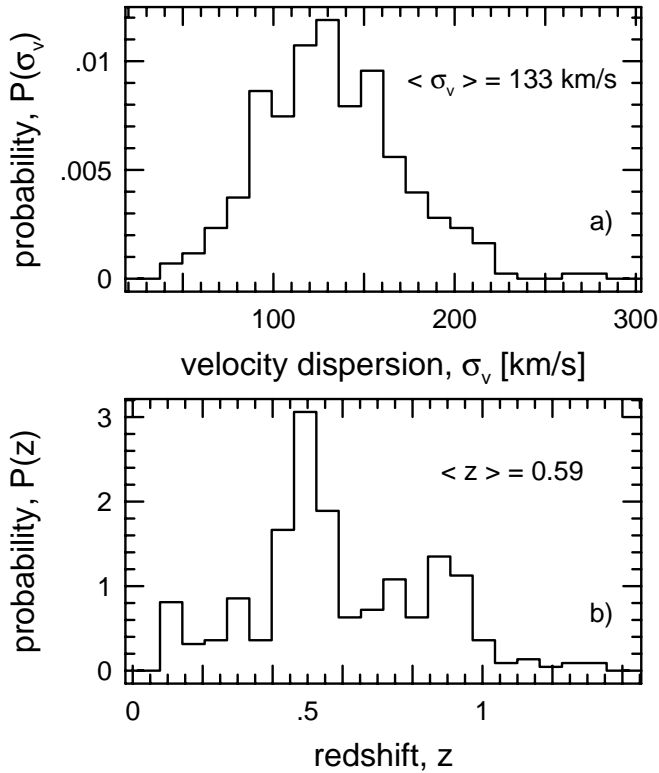}}}
\caption{{\it Top:} Probability distribution for the
velocity dispersions of
the 348 central lens galaxies.
Mean lens velocity dispersion is
$\left< \sigma_v \right> = 133$~km~sec$^{-1}$. {\it Bottom:} Probability 
distributions for the 
redshifts of the
348 central lens galaxies.  Mean lens redshift is $\left< z \right> = 0.59$.}
\label{fig:centrals}
\end{figure}

\begin{figure}[t!]
\figurenum{2}
\centerline{\scalebox{0.75}{\includegraphics{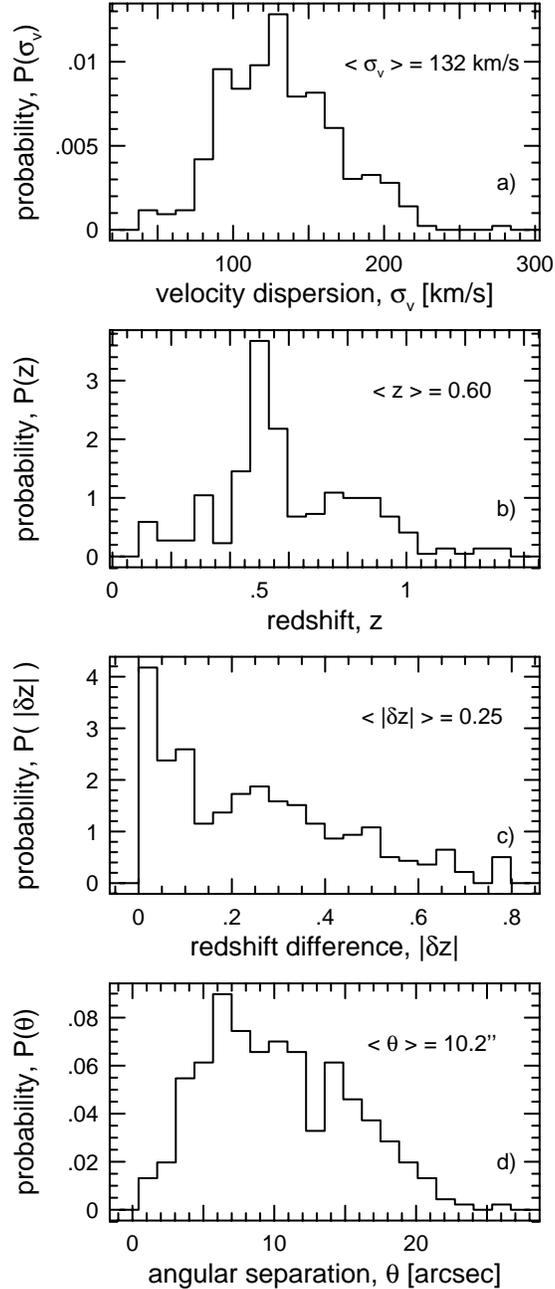}}}
\caption{Probability distributions for the nearest neighbors
of the central lens galaxies.  Panels show: a) velocity dispersion,
b) redshift, c) absolute value of the redshift difference between
the central lens and neighbor, and d) angular separation between the central lens
and neighbor.  Mean values for each of the distributions are indicated
in the panels. 
Note that $|\delta z| <  0.005$ for only $\sim 10$\% of the
nearest neighbors.
}
\label{fig:neighbor_1}
\end{figure}

\begin{figure}[ht!]
\figurenum{3}
\centerline{\scalebox{0.75}{\includegraphics{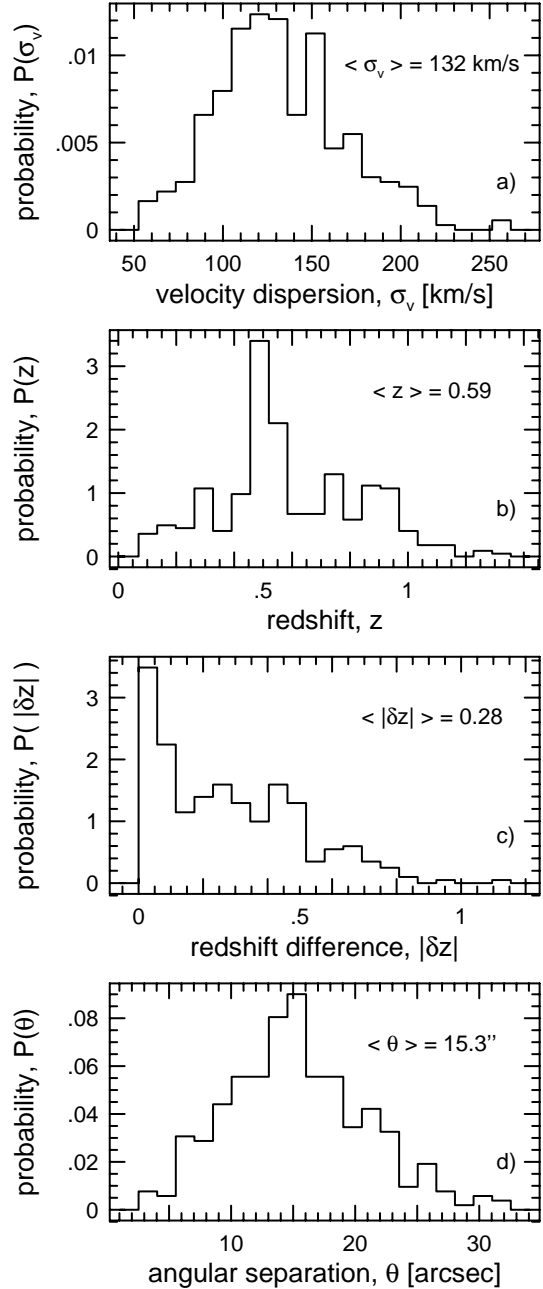}}}
\caption{
Same as Figure~\ref{fig:neighbor_1}, but for the second nearest neighbors.
Note that $|\delta z| < 0.005$ for only $\sim 6$\% of the second nearest neighbors.
}
\label{fig:neighbor_2}
\end{figure}

\begin{figure}[ht!]
\figurenum{4}
\centerline{\scalebox{0.75}{\includegraphics{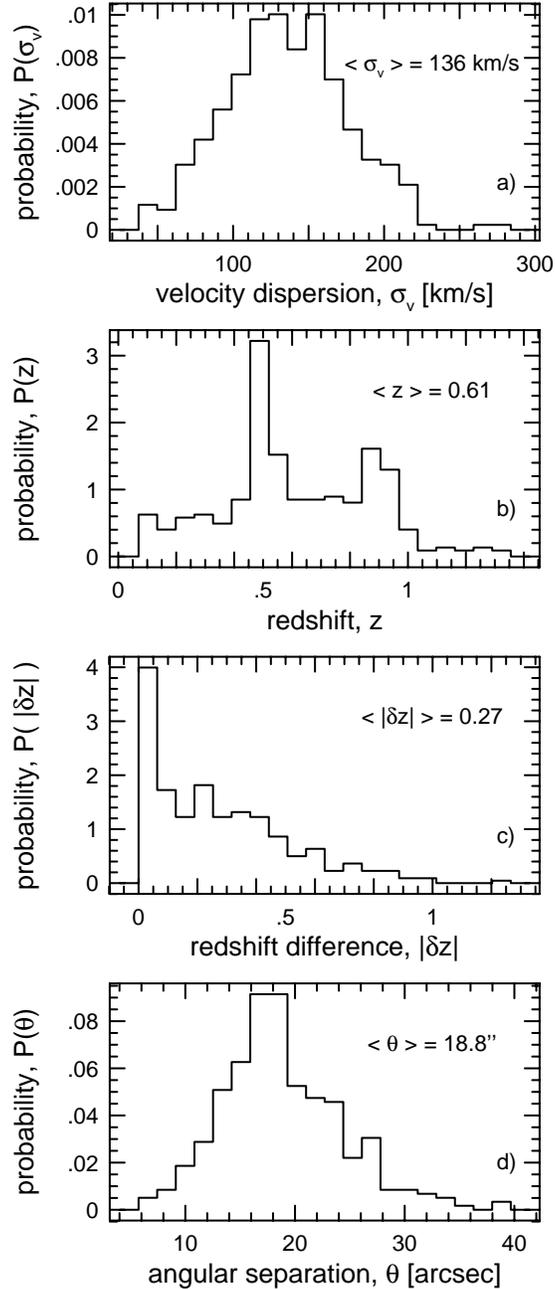}}}
\caption{
Same as Figure~\ref{fig:neighbor_1}, but for the third nearest neighbors.
Note that $|\delta z| < 0.005$ for only $\sim 6$\% of the third nearest neighbors.
}
\label{fig:neighbor_3}
\end{figure}

\begin{figure}[ht!]
\figurenum{5}
\centerline{\scalebox{0.75}{\includegraphics{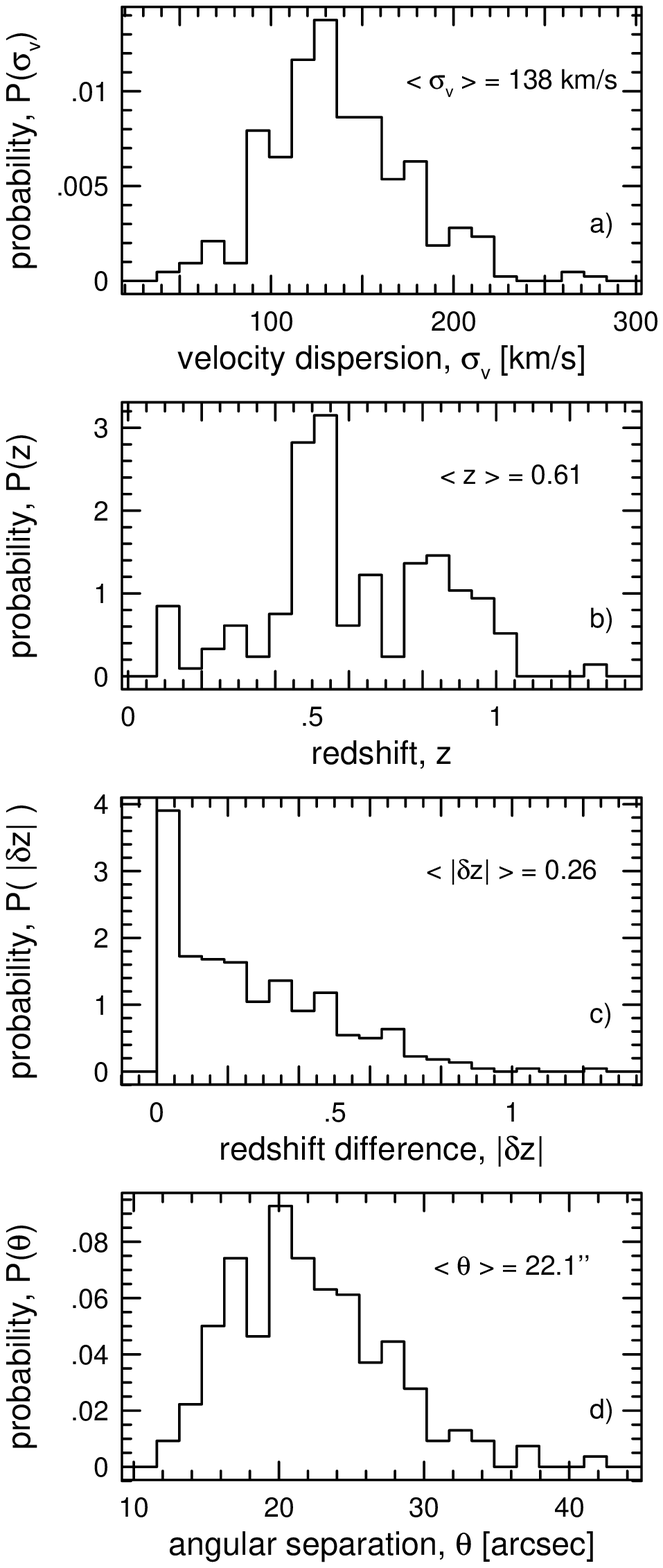}}}
\caption{
Same as Figure~\ref{fig:neighbor_1}, but for the fourth nearest neighbors.
Note that $|\delta z| < 0.005$ for only $\sim 6$\% of the fourth nearest neighbors.
}
\label{fig:neighbor_4}
\end{figure}

The distribution of velocity dispersions for the halos of the 348 central
lens galaxies is shown in the top panel of Figure~\ref{fig:centrals}.  The
bottom panel of Figure~\ref{fig:centrals} shows the redshift distribution for
the 348 central lens galaxies. From Figure \ref{fig:centrals}, the mean halo
velocity dispersion for the central lenses is $\left< \sigma_v \right> = 
133$~km~sec$^{-1}$ and the
mean redshift is $\left< z \right> = 0.59$.  The panels in
Figure~\ref{fig:neighbor_1} show the distributions of a) halo velocity dispersion,
b) redshift, c) absolute redshift difference relative to the central lens, and 
d) angular
separation between the central lens and the galaxies that constitute the ``nearest
neighbor'' to each of the 348 central lenses (i.e., those neighboring 
galaxies with $\theta = \theta_1$).  Figures~\ref{fig:neighbor_2},
\ref{fig:neighbor_3}, and \ref{fig:neighbor_4} show the same distributions
as Figure~\ref{fig:neighbor_1}, but for the second, third, and
fourth nearest neighbors, respectively.  From Figures~\ref{fig:neighbor_1}
through \ref{fig:neighbor_4}, it is clear that the distributions of halo
velocity dispersions and redshifts for the neighboring 
galaxies are nearly identical to those of the 
central lenses.  It is also
clear from Figures~\ref{fig:neighbor_1} through \ref{fig:neighbor_4}
that the distribution of redshift differences between the central lenses
and their near neighbors is similar in all cases. Note that
$\gtrsim 90$\% of the near neighbor galaxies have
$|\delta z| > 0.005$ and are therefore unlikely to be physically associated with the
central lenses.

To compute the mean tangential shear around each of the central lenses,
a fixed source redshift of $z_s = 1.5$ was adopted.  Again, the primary
focus of this investigation
is the net effect of near neighbors on the galaxy-galaxy
lensing signal, so the adoption of a fixed source redshift is sufficient.
The sources were assigned intrinsically 
round shapes and their final shapes were computed as
\begin{equation}
\vec{\chi}_f = \sum_{i=1}^{N_{\rm lens}} \vec{\gamma}_i = \epsilon e^{2i\phi}
\label{eq:chi_f}
\end{equation}
where $\vec{\gamma}_i$ is the shear due foreground lens $i$,
$\epsilon$ is the final, lensed image shape, and $\phi$ is the orientation
of the final, lensed image.
Since all of the simulations were performed in the weak lensing regime, 
Equation {\ref{eq:chi_f} yields the net shear experienced
by each source.  
Simulations were performed separately using only the single nearest 
neighbor, the two nearest neighbors, the three nearest neighbors, and the
four nearest neighbors.  Therefore, the value of $N_{\rm lens}$ in 
Equation (\ref{eq:chi_f}) ranges from 2 to 5.
To compute the galaxy-galaxy lensing signal at a given physical separation, $r_p$,
from a central lens galaxy, a ring of $10^6$ sources, uniformly distributed
in polar angle, was placed around 
the central lens galaxy at a radius of $r_p$ and the final image shapes of
all sources were computed using Equation (\ref{eq:chi_f}) above.  The 
mean tangential shear, centered on the central lens, was then computed in
the standard way (see, e.g., Schneider 2006) using 
the final image shapes of the lensed sources. 

In addition to the mean tangential shear, the excess surface mass density,
$\Delta\Sigma(r_p)$, 
was computed by numerically sampling the mass density field.  
To compute the value of $\Sigma(r_p)$ at the location of a given central lens galaxy,
a ring of $10^6$ sampling points, uniformly distributed
in polar angle, was placed at a radius $r_p$ from the central lens.  
Using Equation (\ref{eq:sigma_sis})
above, the total surface mass
density due to all lenses was computed at each point on the ring, then the mean value
was computed from all $10^6$ sampling points.  To compute the value
of $\left< \Sigma(< r_p) \right>$ at the location of a central lens galaxy, 
$10^6$ sampling points were uniformly distributed within
a circle of radius $r_p$, centered on the central lens.  
Again using Equation (\ref{eq:sigma_sis}), the total
surface mass density due to all lenses was computed at each point within the circle
and the mean value was computed from all $10^6$ sampling points.  
The value of $\Delta\Sigma(r_p)$ was then computed as the difference between
$\left< \Sigma(< r_p) \right>$ obtained from the circle of $10^6$ sampling points and the
value of $\Sigma(r_p)$ obtained from the ring of $10^6$ sampling points.
The sampling
procedure was repeated 100 times at the location of each central lens galaxy 
and the final value of
$\Delta\Sigma(r_p)$ was taken to be the mean of the results from the 100 independent
samplings.

\section{Results} 

Although the spectroscopic
redshifts of the lens galaxies are known, the first set of simulations
that was performed adopted a fixed lens redshift value of $z_l = 0.6$ (i.e., the
mean redshift for the lens galaxies).
The reason for this is to demonstrate directly that, as expected, when 
neighboring lens galaxies are located near to each other in redshift space (i.e., they
are physically related),
the relationship between the mean tangential shear
and the excess surface mass density, given by
\begin{equation}
\gamma_T(r_p) = \Delta\Sigma(r_p) \times \Sigma_c^{-1}, 
\label{eq:shear_esmd}
\end{equation}
is indeed valid.  Shown in Figure~\ref{fig:shear_zfix} is the mean tangential
shear, $\gamma_T(r_p)$, computed around the 348 central lens galaxies 
with fixed redshift $z_l = 0.6$ (open points).
Also shown in Figure~\ref{fig:shear_zfix} is the mean value of the excess surface mass
density, $\Delta\Sigma(r_p)$, scaled by the value of the
critical surface mass density for a 
lens with $z_l = 0.6$ and a source with $z_s = 1.5$ (solid points).  
Error bars for the mean values of $\gamma_T(r_p)$ and $\Delta\Sigma(r_p) \times 
\Sigma_c^{-1}$ were computed using jackknife resampling and in all cases the error
bars are significantly smaller than the sizes of the data points in Figure~\ref{fig:shear_zfix}.
Figure~\ref{fig:shear_zfix} shows that on small scales, the effects of the 
near neighbor galaxies on the galaxy-galaxy lensing is negligible.  On large scales, however, the
effects of weak lensing by the near neighbors are significant and, importantly, the
contributions of the near neighbors to the galaxy-galaxy lensing signal
do not cancel one another.  
Shown in
Figure~\ref{fig:ratio_zfix} is the ratio of the mean tangential shear to the
scaled excess surface mass density.
From Figure~\ref{fig:ratio_zfix}, then,
the ratio of the mean tangential shear to the scaled excess surface mass density is consistent
with unity on all scales.  The error bars in Figure~\ref{fig:ratio_zfix} were computed
by combining the jackknife errors from Figure~\ref{fig:shear_zfix} in quadrature.

\begin{figure}[t!]
\figurenum{6}
\centerline{\scalebox{0.60}{\includegraphics{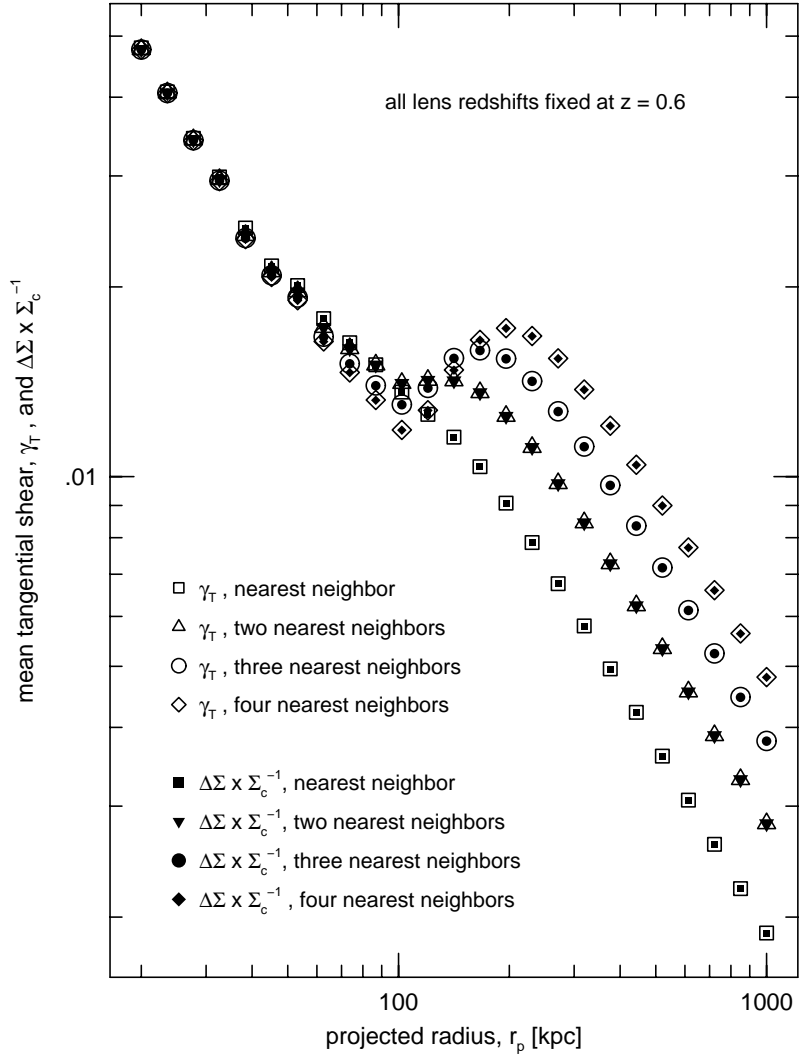}}}
\caption{
Mean tangential shear, $\gamma_T(r_p)$, and scaled excess surface mass density,
$\Delta\Sigma(r_p) \times \Sigma_c^{-1}$,
computed using 
all 348 central lens galaxies, but with the redshifts of the lenses fixed at $z_l = 0.6$.
Squares: central lens and its nearest neighbor only. Triangles: central lens
and its two nearest neighbors.  Circles: central lens and its three nearest neighbors.
Diamonds: central lens and its four nearest neighbors.  Open points: 
$\gamma_T(r_p)$.  Filled points: $\Delta\Sigma(r_p) \times \Sigma_c^{-1}$.
}
\label{fig:shear_zfix}
\end{figure}

\begin{figure}[ht!]
\figurenum{7}
\centerline{\scalebox{0.75}{\includegraphics{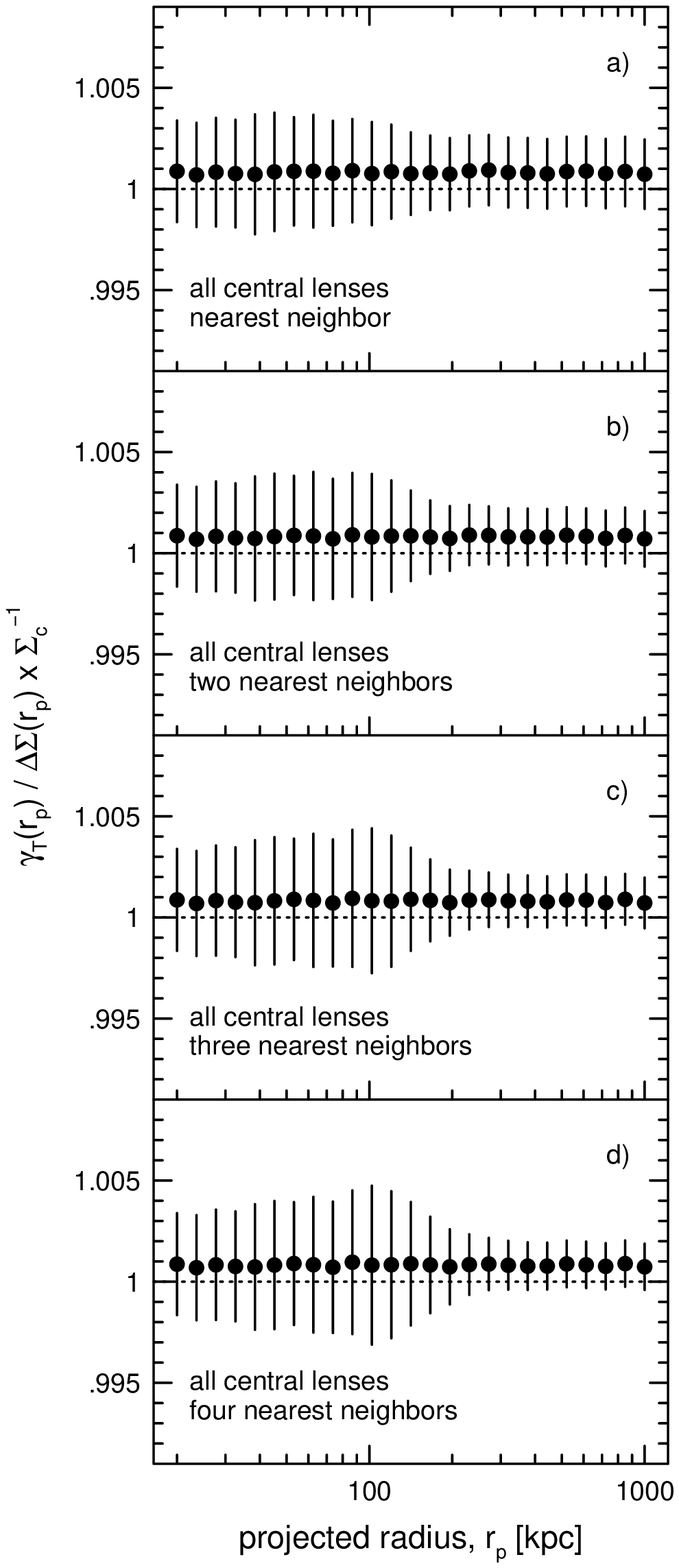}}}
\caption{
Ratios of the curves shown in Figure~\ref{fig:shear_zfix},
$\gamma_T(r_p) \div
\Delta\Sigma(r_p) \times \Sigma_c^{-1}$.
Panels show ratios for different numbers of near neighbors in the
calculations: a) central lens and its nearest neighbor only, b) central lens
and its two nearest neighbors, c) central lens and its three nearest neighbors,
and d) central lens and its four nearest neighbors.  As expected, the ratio
is consistent with unity on all scales.
}
\label{fig:ratio_zfix}
\end{figure}

Having demonstrated the validity of Equation (\ref{eq:shear_esmd}) for the special
limit in which the central lens and its near neighbors have identical 
redshifts, a second set of simulations was run and in this second
set of simulations the galaxies were assigned their
observed spectroscopic redshifts.  Here, when the value of $\Sigma_c$ is taken to be
the critical surface mass density of the central lens, the validity of 
Equation (\ref{eq:shear_esmd}) is no longer assured since the
near neighbor galaxies are typically located at redshifts that are significantly
different from that of the central lens (see panel~c in 
Figures~\ref{fig:neighbor_1} through \ref{fig:neighbor_4}).   Shown in 
Figure~\ref{fig:outliers} are two different extreme examples of the failure of
Equation (\ref{eq:shear_esmd}) when a central lens galaxy and its nearest
neighbor are located at significantly different redshifts.  In Figure~\ref{fig:outliers}
the calculations were performed using the central
lens galaxy and its single nearest neighbor.  In both cases, the halo of the
central lens galaxy in Figure~\ref{fig:outliers} has
a velocity dispersion that is comparable to the mean value for all 348 central
lenses.  In the top panel of Figure~\ref{fig:outliers}, $\gamma_T(r_p)$ agrees with
$\Delta\Sigma(r_p) \times \Sigma_c^{-1}$ on small scales; however, on large scales 
$\Delta\Sigma(r_p) \times \Sigma_c^{-1}$ exceeds
$\gamma_T(r_p)$ by $\sim 30$\%.  (Note
that the steep ``dip'' in the curves shown in the top panel of Figure~\ref{fig:outliers}
occurs at a scale comparable to the spacing between the two lenses and such features
are commonly seen in theoretical weak lensing profiles when more than one lens
is present.)  In the bottom panel of Figure~\ref{fig:outliers}, $\gamma_T(r_p)$
exceeds $\Delta\Sigma(r_p)
\times \Sigma_c^{-1}$ by $\sim 30$\% over all but the smallest
scales that are shown.  
The velocity dispersions of the two lenses in the top panel are
132~km~sec$^{-1}$ (central lens) and 214~km~sec$^{-1}$ (nearest neighbor).  The
redshifts of the two lenses in the top panel are 0.85 (central lens) and 1.13 (nearest
neighbor), and the separation between the lenses is $\theta_1 = 8.9"$ (corresponding to 
a projected separation of $r_p \sim 70$~kpc at the redshift of the central lens).
The velocity dispersions of the two lenses in the bottom panel are
134~km~sec$^{-1}$ (central lens) and 128~km~sec$^{-1}$ (nearest neighbor).  The
redshifts of the two lenses in the bottom panel are 0.93 (central lens) and 0.52 (nearest
neighbor), and the separation between the lenses is $\theta_1 = 2.2"$ (corresponding to 
a projected separation of $r_p \sim 18$~kpc at the redshift of the central lens).

\begin{figure}[ht!]
\figurenum{8}
\centerline{\scalebox{0.75}{\includegraphics{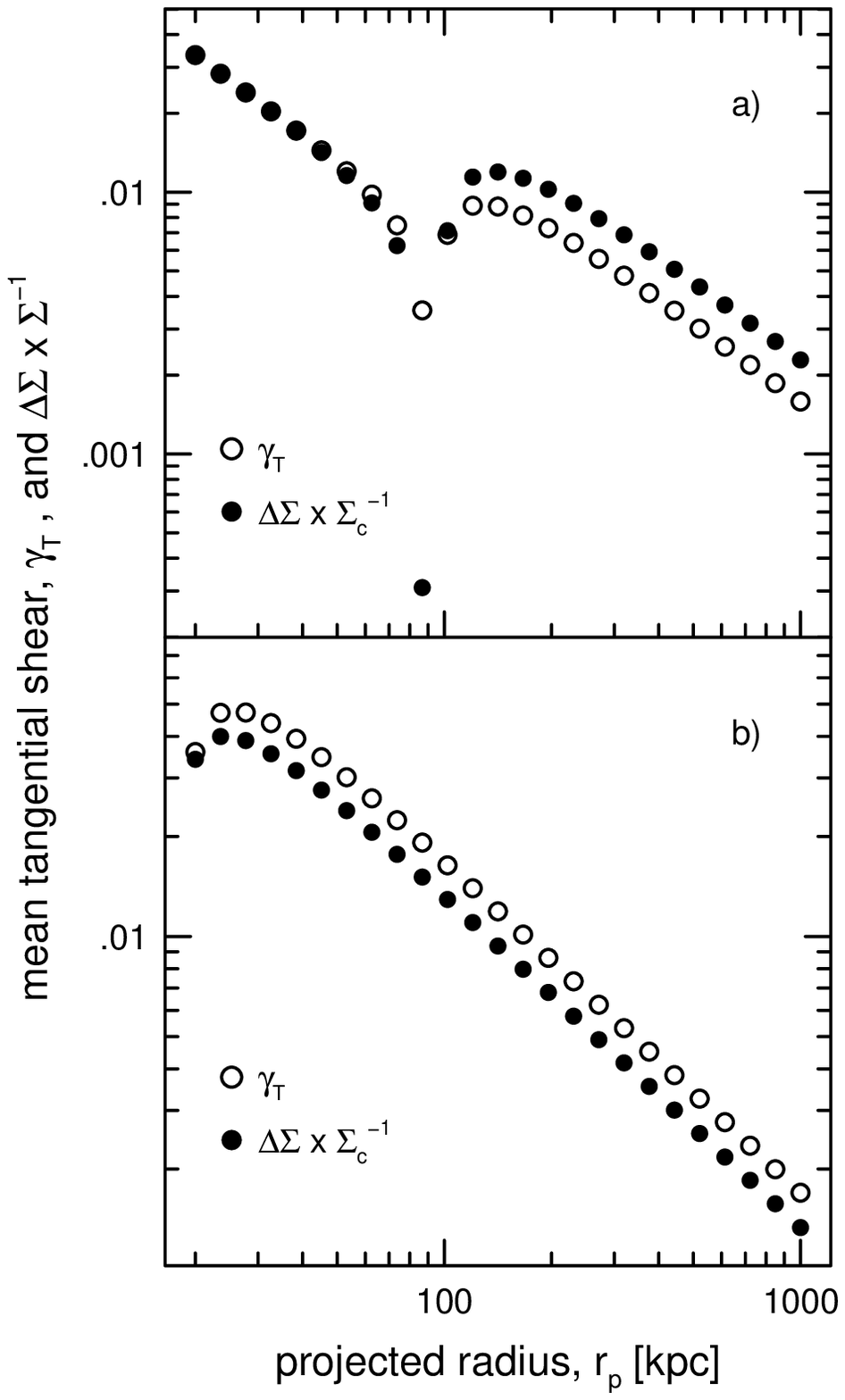}}}
\caption{
Two extreme examples of the failure of Equation~(\ref{eq:shear_esmd}) when
the central lens galaxy and the nearest neighbor lens galaxy have different redshifts.
{\it Top:} Central lens has velocity dispersion $\sigma_v = 132$~km~sec$^{-1}$ and
redshift $z_l = 0.85$.  Nearest neighbor lens has velocity dispersion
$\sigma_v = 214$~km~sec$^{-1}$ and redshift $z_l = 1.13$.  Separation between the
two lenses on the sky corresponds to $r_p \sim 70$~kpc at the redshift of the central lens.
{\it Bottom:} Central lens has velocity dispersion $\sigma_v = 134$~km~sec$^{-1}$ and
redshift $z_l = 0.93$.  Nearest neighbor lens has velocity dispersion
$\sigma_v = 128$~km~sec$^{-1}$ and redshift $z_l = 0.52$.  Separation between the
two lenses on the sky corresponds to $r_p \sim 18$~kpc at the redshift of the central lens.
Circles: mean tangential shear, $\gamma_T(r_p)$.  Squares: scaled excess surface
mass density,
$\Delta\Sigma(r_p) \times \Sigma_c^{-1}$, where $\Sigma_c$ is the value of the
critical surface mass density for the central lens.
}
\label{fig:outliers}
\end{figure}

Although the differences between $\gamma_T(r_p)$ and $\Delta\Sigma(r_p) \times
\Sigma_c^{-1}$ are substantial for the two central lenses shown in
Figure~\ref{fig:outliers}, the sense of the discrepancy (i.e., greater or less than unity) is
opposite.  Therefore, it is to be expected that, when computed
over the entire population of central lens galaxies, some of the discrepancy
between $\gamma_T(\theta)$ and $\Delta\Sigma(r_p) \times \Sigma_c^{-1}$ 
that is seen for individual lenses will necessarily be averaged away.
Results for the mean tangential shear, averaged over all 348 central lenses,
are shown by the open points in 
Figure~\ref{fig:shear_all}.  Solid points in Figure~\ref{fig:shear_all} show the
mean value of $\Delta\Sigma(r_p) \times \Sigma_c^{-1}$, where $\Sigma_c$ is the
value of the critical surface mass density that is appropriate for the central
lens, given its actual redshift.  
This is analogous to observational studies of galaxy-galaxy lensing in which
the observed mean tangential shear is converted into a value of the excess
surface mass density by multiplying by the critical surface mass density for the
central lens.  Here jackknife error bars were again
computed and all of the resulting error bars are smaller than the individual
data points in Figure~\ref{fig:shear_all}.  As with the case when all lens galaxies
were placed at a fixed redshift of $z_l = 0.6$ (Figure~\ref{fig:shear_zfix} above), 
the effect of the near neighbor galaxies on the mean tangential shear
is small on small scales, but is significant
on large scales. Again, as in Figure~\ref{fig:shear_zfix},
the effects of the near neighbor galaxies on the
mean tangential shear do not cancel each other on large scales.  Also, unlike the case when all
lens galaxies were placed a single, fixed redshift, a difference between
the values of $\gamma_T(r_p)$ and $\Delta\Sigma(r_p) \times \Sigma_c^{-1}$ manifests
when the redshifts of the lens galaxies are taken to be their actual, observed
spectroscopic redshifts.  That is, Equation (\ref{eq:shear_esmd}) is no longer strictly
valid due to the fact that the value of $\Sigma_c$  necessarily differs for the central
lenses and their near neighbors.

\begin{figure}[ht!]
\figurenum{9}
\centerline{\scalebox{0.60}{\includegraphics{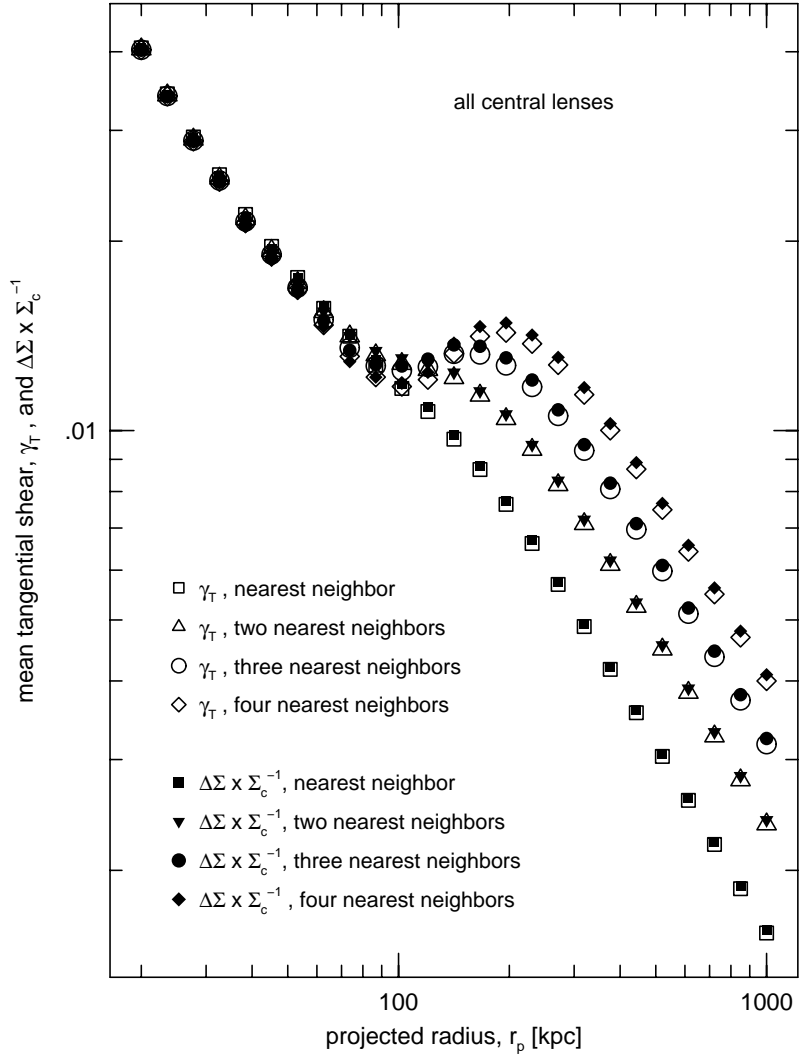}}}
\caption{
Mean tangential shear, $\gamma_T(r_p)$, and scaled excess surface mass density,
$\Delta\Sigma(r_p) \times \Sigma_c^{-1}$,
computed using 
all 348 central lens galaxies. Unlike Figure~\ref{fig:shear_zfix}, 
here the redshifts of the lenses are taken
to be their observed spectroscopic redshifts.
Squares: central lens and its nearest neighbor only. Triangles: central lens
and its two nearest neighbors.  Circles: central lens and its three nearest neighbors.
Diamonds: central lens and its four nearest neighbors.  Open points: 
$\gamma_T(r_p)$.  Filled points: $\Delta\Sigma(r_p) \times \Sigma_c^{-1}$.
}
\label{fig:shear_all}
\end{figure}

Shown in Figure~\ref{fig:ratio_all} is the ratio of the mean tangential shear
to the scaled excess surface mass density
for the curves shown in Figure~\ref{fig:shear_all}.
As with Figure~\ref{fig:ratio_zfix}, the error bars in Figure~\ref{fig:ratio_all} were
obtained by combining the error bars from Figure~\ref{fig:shear_all} in quadrature.
From Figure~\ref{fig:ratio_all}, it is clear that, over the majority of scales for
which the weak lensing signal was computed, the value of $\Delta\Sigma(r_p) \times
\Sigma_c^{-1}$ exceeds the value of $\gamma_T(r_p)$.
The ratio of $\gamma_T(r_p)$ to $\Delta\Sigma(r_p) \times \Sigma_c^{-1}$
is not monotonic and ranges from unity on the smallest scales 
($r_p \lesssim 20$~kpc) to
greater than unity on larger, but ``galactic'', scales ($30~{\rm kpc} \lesssim r_p
\lesssim 60$~kpc) to less than unity on the largest scales ($r_p \gtrsim 100$~kpc).
From Figures~\ref{fig:shear_all} and \ref{fig:ratio_all}, it is also clear that
adding an increasing number of near neighbors into the calculation 
of the net weak lensing signal
increases the value of the mean tangential shear at large lens-source
separations. Adding an increasing number near neighbors also increases
the degree to which $\gamma_T(r_p)$ and $\Delta\Sigma(r_p) \times \Sigma_c^{-1}$ 
disagree with each other on the largest scales (i.e., ranging from a difference of $\sim 1$\% for
the case of one central lens and its single nearest neighbor to a difference of
$\sim 2$\% for the case of one central lens and its four nearest neighbors).

\begin{figure}[ht!]
\figurenum{10}
\centerline{\scalebox{0.75}{\includegraphics{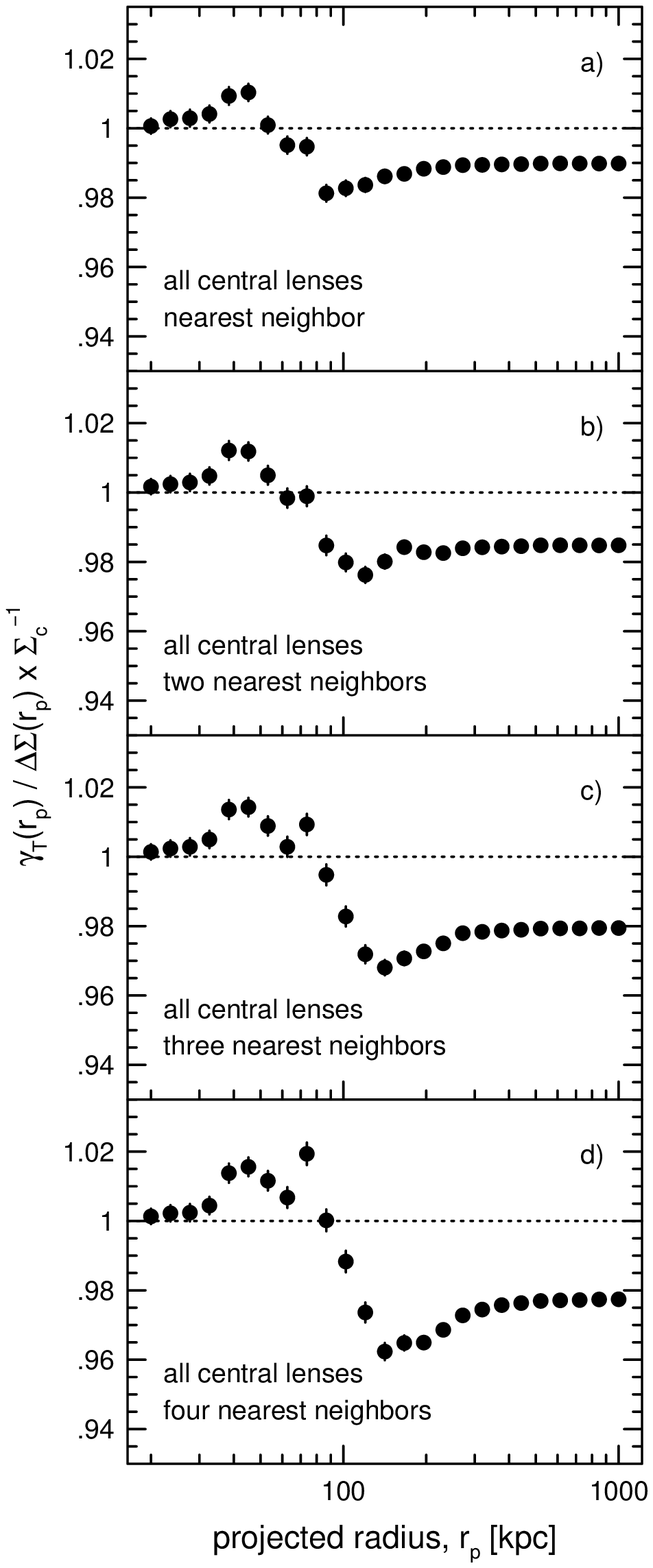}}}
\caption{
Ratios of the curves shown in Figure~\ref{fig:shear_all},
$\gamma_T(r_p) \div
\Delta\Sigma(r_p) \times \Sigma_c^{-1}$.
Panels show ratios for different numbers of near neighbors:
a) central lens and its nearest neighbor only, b) central lens
and its two nearest neighbors, c) central lens and its three nearest neighbors,
and d) central lens and its four nearest neighbors.  
Unlike Figure~\ref{fig:ratio_zfix}, here the redshifts of the lenses
are taken to be their observed spectroscopic redshifts.  The broad redshift
distribution of the lens galaxies causes the ratio of the mean tangential
shear to the scaled excess surface mass density to deviate from
unity over most scales.
}
\label{fig:ratio_all}
\end{figure}

The results shown in Figures~\ref{fig:shear_all} and \ref{fig:ratio_all} are those
that are obtained when $\gamma_T(r_p)$ and $\Delta\Sigma(r_p) \times \Sigma_c^{-1}$
are averaged over the entire population of 348 central lenses.  Given the
wide range of velocity dispersions for the halos of the lens galaxies, however, it is entirely 
possible that the net effects of the near neighbors on the weak
lensing signal could vary with the velocity dispersions of the central lenses.
To investigate this, $\gamma_T(r_p)$ and $\Delta\Sigma(r_p)
\times \Sigma_c^{-1}$ were separately computed using central lenses
with higher than average halo
velocity dispersions ($\sigma_v > 133$~km~sec$^{-1}$, 166 galaxies) and
lower than average halo velocity dispersions
($\sigma_v < 133$~km~sec$^{-1}$, 182 galaxies).  Results for central lenses
with higher than average velocity dispersions are shown in Figures~\ref{fig:shear_high}
and \ref{fig:ratio_high}.  Results for central lenses with lower than
average velocity dispersions are shown in Figures~\ref{fig:shear_low} and
\ref{fig:ratio_low}.
Over the entire range of scales for which the signal was computed, the ratio
of $\gamma_T(r_p)$ to $\Delta\Sigma(r_p) \times \Sigma_c^{-1}$ for the high-velocity
dispersion central lenses tracks in the opposite direction than it does for the
low-velocity dispersion central lenses.  That is, when the ratio is greater than unity
for the high-velocity dispersion central lenses, it is less than unity for the low-velocity
dispersion central lenses, and when the ratio is less than unity for the high-velocity
dispersion central lenses it is greater than unity for the low-velocity dispersion 
central lenses.  Because of this, the differences between $\gamma_T(r_p)$ and
$\Delta\Sigma(r_p) \times \Sigma_c^{-1}$ when averaged over the entire 
population of lenses are significantly less than when they are averaged over the subsets of
high- and low-velocity dispersion central lenses.

\begin{figure}[ht!]
\figurenum{11}
\centerline{\scalebox{0.60}{\includegraphics{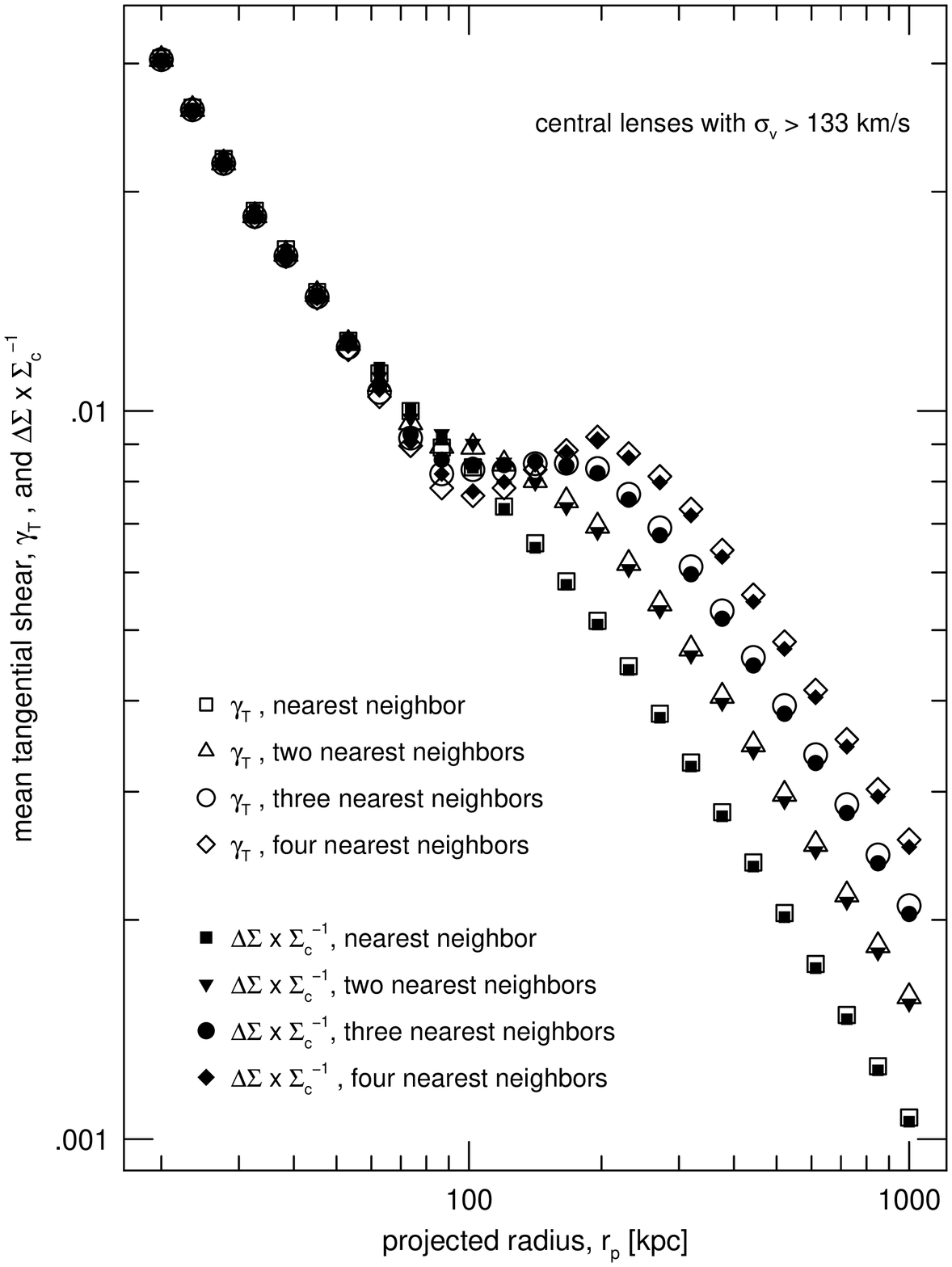}}}
\caption{
Same as Figure~\ref{fig:shear_all}, but for the 166 central lenses 
with higher than average velocity dispersions.
}
\label{fig:shear_high}
\end{figure}

\begin{figure}[ht!]
\figurenum{12}
\centerline{\scalebox{0.60}{\includegraphics{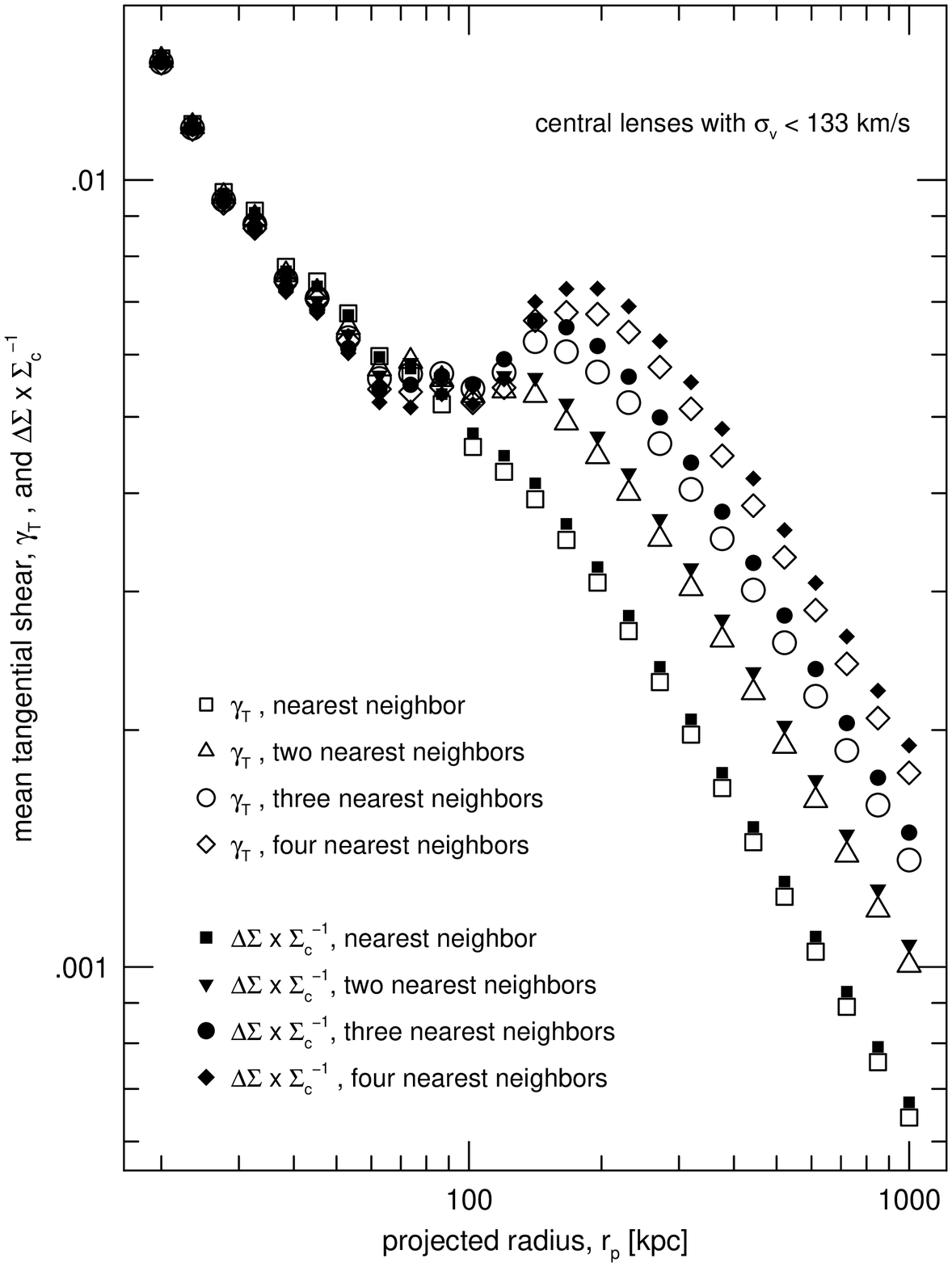}}}
\caption{
Same as Figure~\ref{fig:shear_all}, but for the 182 central lenses 
with lower than average velocity dispersions.
}
\label{fig:shear_low}
\end{figure}

\begin{figure}[ht!]
\figurenum{13}
\centerline{\scalebox{0.75}{\includegraphics{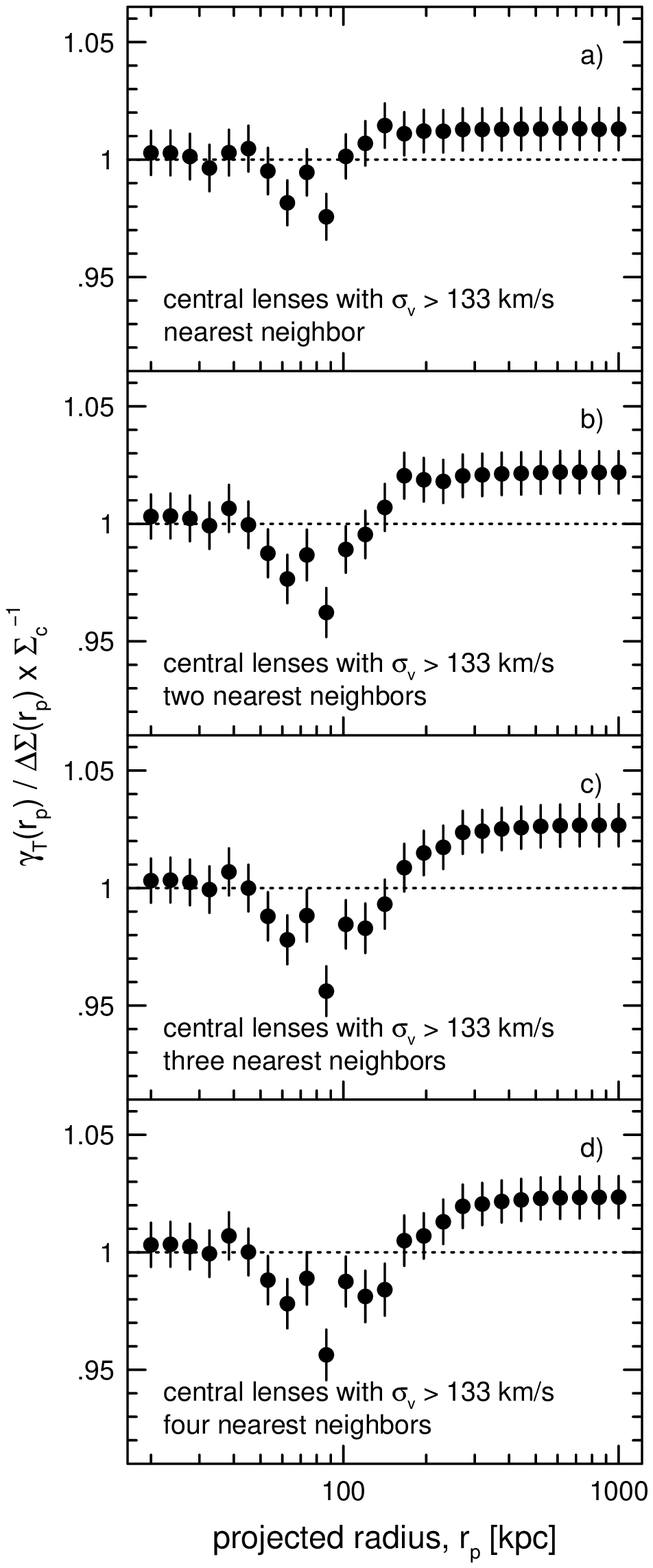}}}
\caption{
Same as Figure~\ref{fig:ratio_all}, but for the 166 central lenses 
with higher than average velocity dispersions.
}
\label{fig:ratio_high}
\end{figure}

\begin{figure}[ht!]
\figurenum{14}
\centerline{\scalebox{0.75}{\includegraphics{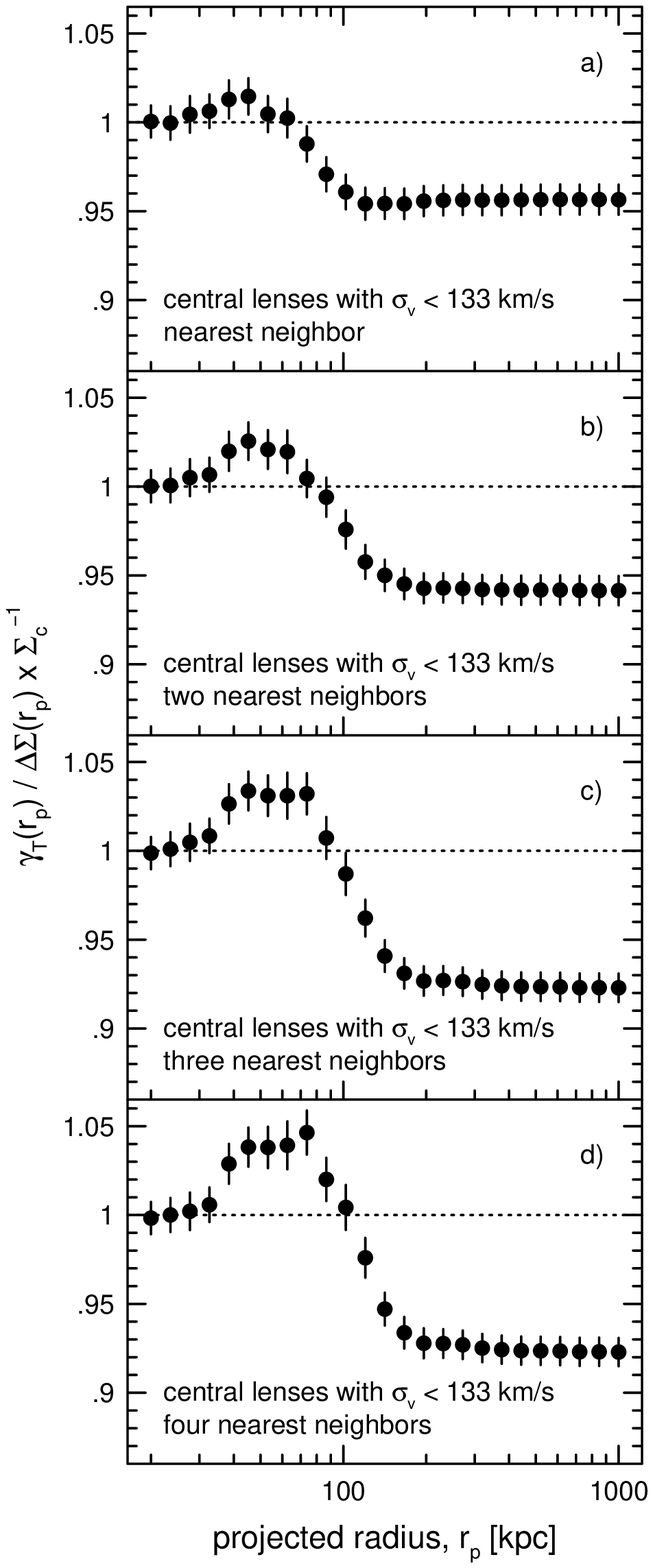}}}
\caption{
Same as Figure~\ref{fig:ratio_all}, but for the 182 central lenses 
with lower than average velocity dispersions.
}
\label{fig:ratio_low}
\end{figure}

From Figures~\ref{fig:shear_high}, \ref{fig:shear_low}, \ref{fig:ratio_high}, and
\ref{fig:ratio_low}, it is clear that physically unrelated
near neighbors affect the 
weak lensing signal for both high- and low-velocity dispersion central 
lenses, but to differing degrees. In the case of high-velocity
dispersion central lenses
(Figures~\ref{fig:shear_high} and \ref{fig:ratio_high}), 
the value of $\gamma_T(r_p)$ on large scales
exceeds that of $\Delta\Sigma(r_p) \times \Sigma_c^{-1}$,
and the difference between the two quantities ranges from $\sim 1.5$\% for
the case of one central lens and its single nearest neighbor to $\sim 2.5$\%
for the case of one central lens and its four nearest neighbors.  In the case
of low-velocity dispersion central lenses
(Figures~\ref{fig:shear_low} and \ref{fig:ratio_low}), the weak
lensing signal is
much more strongly affected by physically unrelated near neighbors than
is the weak lensing signal for high-velocity dispersion central lenses.
On the largest scales, the differences between $\gamma_T(r_p)$ and
$\Delta\Sigma(r_p) \times \Sigma_c^{-1}$ for the low-velocity dispersion
central lenses range from $\sim 4.5$\% (for the case of a central lens and its
single nearest neighbor) to $\sim 7$\% (for the case of a central lens and its
four nearest neighbors).

Lastly, shown in Figure~\ref{fig:ratio_dist} is the probability distribution
for the ratio of the mean tangential shear to the scaled excess surface
mass density, evaluated at $r_p = 1$~Mpc.  The
left panels in Figure~\ref{fig:ratio_dist} show the results averaged over all 
348 central lens galaxies, while the center and right panels show the results 
for the low-velocity dispersion central lenses and the
high-velocity dispersion central lenses, respectively.  From top to bottom,
the panels in Figure~\ref{fig:ratio_dist} show the results for calculations
that include only the nearest neighbor (topmost panels), the two nearest neighbors,
the three nearest neighbors, and the four nearest neighbors (bottommost
panels).  The
dotted vertical lines indicate a value of unity.  From
Figure~\ref{fig:ratio_dist}, then, in all cases the ratio of
the mean tangential shear to the scaled excess surface mass density at large scales
has a probability distribution that is both broad and asymmetrical.  
For a given central lens, the ratio of the mean tangential shear to 
the scaled excess surface mass density at $r_p = 1$~Mpc
spans a range from $\sim 0.5$ to $\sim 1.5$,
and the shape of the probability distribution is a strong function of the 
velocity dispersions of the central lenses.  That is, for a given
low-velocity dispersion central lens, the asymmetry of the probability 
distribution is such that the scaled excess surface mass density typically
exceeds the mean tangential shear.  For a given high-velocity dispersion
central lens, the asymmetry of the probability distribution is such that the
mean tangential shear tends to exceed the scaled excess surface mass density.

\begin{figure}[ht!]
\figurenum{15}
\centerline{\scalebox{0.60}{\includegraphics{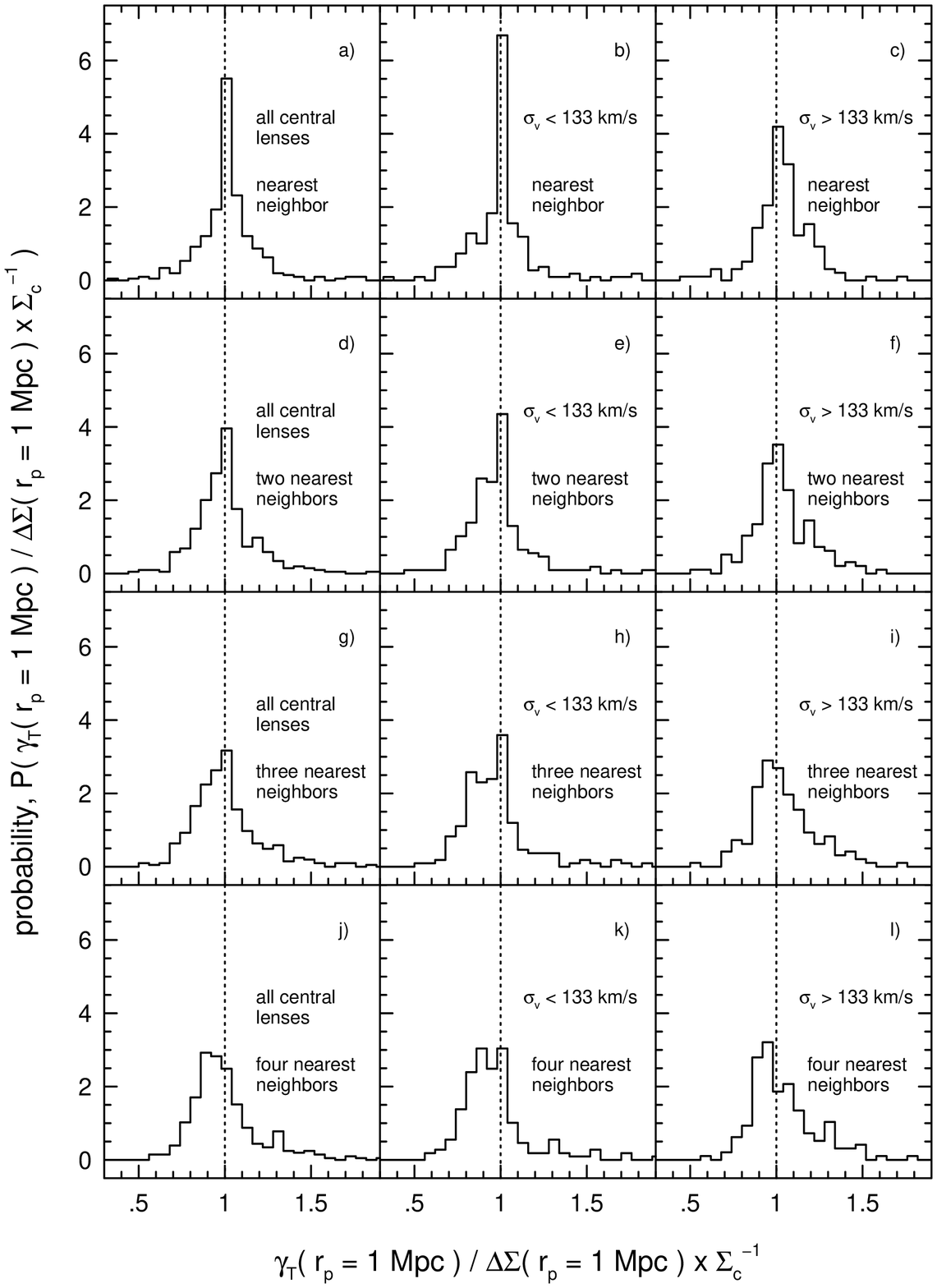}}}
\caption{
Probability distributions for the ratio of the mean tangential shear
to the scaled excess surface mass density, evaluated at a projected
radial distance of $r_p = 1$~Mpc from the central lenses.  Vertical
dotted lines indicate a value of unity.  {\it Left:}
Results from all 348 central lenses.  {\it Center:} Results from 
the 182 central lenses with lower than average velocity dispersions.
{\it Right:} Results from the 168 central lenses with higher than
average velocity dispersions.  From top to bottom, the panels show the results
for the central lens and its nearest neighbor (topmost panels), the central
lens and its two nearest neighbors, the central lens and its three
nearest neighbors, and the central lens and its four nearest neighbors 
(bottommost panels).
For a given central lens, the ratio takes on 
a wide range of values (from $\sim 0.5$ to $\sim 1.5$) and the probability
distribution
is highly asymmetrical in all cases.  Averaged over the population
of central lenses, the asymmetry of the probability 
distribution leads to a net discrepancy between the mean tangential shear
and the scaled surface mass density (see Figures~\ref{fig:ratio_all},
\ref{fig:ratio_high}, and \ref{fig:ratio_low}).
}
\label{fig:ratio_dist}
\end{figure}

\section{Summary and Discussion}

The observed celestial coordinates, spectroscopic redshifts, and
rest-frame blue luminosities of galaxies in the region of the HDF-N were
used as the basis of a suite of Monte Carlo simulations of weak
galaxy-galaxy lensing.  Since the simulations incorporate known galaxies
with a uniform completeness in redshift space ($R \le 23$), the simulations
naturally incorporate the intrinsic redshift distribution, mass distribution,
and clustering of galaxies in our Universe.  The simulations investigated
the effects of near-neighbor galaxies on the mean tangential shear, computed
around central lens galaxies, and the relationship between the
mean tangential shear and the excess surface mass density,
scaled by the critical surface mass density of the central lens. 
The main results from the simulations are as follows:

\begin{enumerate}
\item As expected, the relationship
between the mean tangential shear and the scaled excess surface mass density is given
by $\gamma(r_p) = \Delta\Sigma(r_p) \times \Sigma_c^{-1}$ when all
lens galaxies are assigned a fixed, identical redshift. 

\item When the lens galaxies are assigned their observed spectroscopic redshifts,
the value of $\gamma(r_p)$ may differ from that of $\Delta\Sigma(r_p) \times 
\Sigma_c^{-1}$.  This is due to the fact that the majority of the near neighbors
have redshifts that are significantly different from those
of the central lenses.

\item At large scales, the ratio of 
the mean tangential shear to the scaled excess surface mass density 
for a given central lens galaxy
spans a wide range, from $\sim 0.5$ to $\sim 1.5$.  

\item The magnitude and sense (i.e., greater or less than unity) of the
discrepancy between the mean tangential shear and the scaled excess surface
mass density are functions of 
both the physical scale, $r_p$, and the velocity dispersions of the central lenses.
In particular, at large scales the difference between the mean
tangential shear and the scaled excess surface mass density is considerably greater for
low-velocity dispersion central lenses than it is for high-velocity dispersion
central lenses.

\item The effects of physically unrelated near-neighbors on the 
weak lensing shear signal do not cancel out.  Instead, physically unrelated
near-neighbors increase the mean tangential shear on scales larger than
the separation between the central lens and its neighbors.

\end{enumerate}

The significance of this work is a demonstration that, in the limit of
a realistic redshift distribution for weak galaxy lenses, systematic
errors should arise when the common
method of converting an observed galaxy-galaxy lensing signal into a constraint
on the excess surface mass density 
is used
(i.e., multiplication of the
value of $\gamma_T$ for a given source galaxy by the value of $\Sigma_c$ for
the central lens and that particular source galaxy).  This is because
neighboring lens galaxies 
that are located at redshifts other than the redshift of the central lens
give rise to a substantial component of the net weak lensing shear.  
An important goal for future, deep weak lensing surveys (for example, those that will be
yielded by the Large Synoptic Survey Telescope, {\sl Euclid}, and {\sl WFIRST}) is a
constraint on the mass distribution of the universe that is accurate to 
$\lesssim 1$\% on scales $r_p > 1$~Mpc.  The results of this work suggest, therefore,
that in order to achieve such an accurate measurement of the mass density from future
galaxy-galaxy lensing studies, it may be important to move beyond the methods that
are currently used to convert the observed galaxy-galaxy lensing signal into a 
measurement of the excess surface mass density.  This is due to the fact that, to
date, most theoretical predictions for the galaxy-galaxy lensing signal have
adopted a model in which the two-halo term (i.e., the contribution to the weak
lensing signal caused by galaxies other than the central lens) is
attributed to neighboring galaxies that are physically related to the 
central galaxy (i.e., the neighboring lens galaxies that contribute 
to the net shear are assumed to be located
at the same redshift as the central galaxy).  Noteable exceptions to this
include Schrabback et al.\ (2015) and Saghiha et al.\ (2016), both of which utilized
direct ray-traying through N-body simulations to predict the galaxy-galaxy lensing signal
using the methods described by Hilbert et al.\ (2009).

The simulations presented here
demonstrate that, in a sufficiently deep data set, the majority of near neighbors
are not physically related to the central lens, and these
physically unrelated near neighbors
give rise to a significant component of the net galaxy-galaxy lensing
signal on large scales.
Based upon the results presented here,
the need to explicitly include the effects of physically unrelated near
neighbors
may be especially important when attempting to constrain 
the dependence of $\Delta\Sigma(r_p)$ on the physical properties of the
central lens galaxies (e.g., luminosity, color, stellar mass).  Indeed, although
the significance is somewhat low, there may already be an indication in existing 
galaxy-galaxy lensing observations that physically unassociated near neighbors
are contributing to the observed signal.  For example, Velander
et al.\ (2014) found that their adopted model for the one- and two-halo
terms provided good fits to the galaxy-galaxy lensing signal when the signal
was computed using
the entire population of central lens galaxies and also when it was computed 
separately for 
red central lenses.  However, their model did not provide a good overall fit to the
galaxy-galaxy lensing signal
for blue central lenses
with low luminosities and low stellar masses.  The work presented here suggests
that, if physically unassociated near neighbors are contributing to the galaxy-galaxy
lensing signal, then the largest
discrepancies between the model adopted
by Velander et al.\ (2014) and the observed mean tangential shear should, in fact,
manifest in the regime of the lowest-luminosity, lowest-mass
central galaxies.

The degree to which the observed mean tangential shear will differ from the actual
scaled excess surface mass density in any large, future survey will, of course, depend upon
the redshift distribution of the foreground lenses and the nature of the dark matter
halos that surround them.  For the sake of simplicity, here an unphysical isothermal
sphere model was adopted for the halos of the lens galaxies.  This was done in order
to construct a demonstration of the effects of near neighbor galaxies on the 
galaxy-galaxy lensing signal that is straightforward to reproduce.  Because of the
simplicity of the adopted halo mass distribution, and because the effects of all
possible foreground lens galaxies were not expressly included here (i.e., the
simulations were limited to at most the contribution of the four nearest neighbors),
the simulations presented here do not represent an accurate estimate of the degree
of discrepancy between the mean tangential shear and the scaled excess surface mass
density that would be expected to occur in our Universe. Rather, it is generally 
agreed that the mass density of our Universe is dominated by CDM, for which the
dark matter halos of galaxies are not isothermal.  The shear profile
of a CDM halo 
differs from that of an isothermal sphere, being shallower than
isothermal on 
scales less than the scale radius of the halo and steeper than isothermal on scales
greater than the scale radius of the halo (see, e.g., Figure~1 of Wright \& Brainerd
2000).  An accurate estimate of the degree of discrepancy between the mean
tangential shear and the scaled excess surface mass density 
that could be expected in our Universe therefore
awaits a more thorough analysis, such as could be obtained from
high-resolution CDM simulations that include luminous galaxies via either semi-analytic
galaxy formation or numerical hydrodynamics.

\acknowledgments

Animated, insightful conversations with Kelly Blumenthal, Dave Goldberg, and Brandon Harrison,
as well as support from the NASA Astrophysics Theory Program
via grant NNX13AH24G~S04, are gratefully acknowledged.




\listofchanges

\end{document}